\documentclass[conference]{IEEEtran}
\pdfoutput=1
\usepackage[
    backend=biber,
    sorting=none,
    citestyle=numeric,
    bibstyle=numeric,
]{biblatex}
\usepackage{amsmath,amssymb,amsfonts}
\usepackage{algorithmic}
\usepackage{graphicx}
\usepackage{textcomp}
\usepackage{xcolor}
\usepackage{hyperref}
\usepackage{flushend}
\usepackage[normalem]{ulem}

\title{Hybrid DLT as a data layer for real-time, data-intensive applications}

\author{
    \IEEEauthorblockN{Andrea Canciani}
    \IEEEauthorblockA{
        \textit{Geckosoft}\\
        a.canciani@geckosoft.it
    }
    \and
    \IEEEauthorblockN{Claudio Felicioli}
    \IEEEauthorblockA{
        \textit{Traent}\\
        claudio.felicioli@traent.com
    }
    \and
    \IEEEauthorblockN{Andrea Lisi}
    \IEEEauthorblockA{
        \textit{University of Pisa and}\\ \textit{Consiglio Nazionale}\\ \textit{delle Ricerche - IIT}\\
        andrea.lisi@phd.unipi.it
    }
    \and
    \IEEEauthorblockN{Fabio Severino}
    \IEEEauthorblockA{
        \textit{Traent}\\
        fabio.severino@traent.com
    }
}

\begin{document}

\maketitle

\thispagestyle{plain}
\pagestyle{plain}

\begin{abstract}
    We propose a new approach, termed Hybrid DLT, to address a broad range of industrial use cases where certain properties of both private and public DLTs are valuable, while other properties may be unnecessary or detrimental. The Hybrid DLT approach involves a system where private ledgers, with limited data block dissemination, are collaboratively created by nodes within a private network. The Notary, a publicly auditable authoritative component, maintains a single, official, coherent history for each private ledger without requiring access to data blocks. This is achieved by leveraging a public DLT solution to render the ledger histories tamper-proof, consequently providing tamper-evidence for ledger data disclosed to external actors. We present Traent Hybrid Blockchain, a commercial implementation of the Hybrid DLT approach: a real-time, data-intensive collaboration system for organizations seeking immutable data while also needing to comply with the European General Data Protection Regulation (GDPR).
\end{abstract}


\section{Introduction}

\noindent Blockchain technology was first introduced alongside Bitcoin \cite{Bitcoin} as a Peer-to-Peer (P2P) network, in which peers collaborate to maintain a shared state, despite not knowing or trusting one another. In the case of Bitcoin, the shared state is a ledger containing all user transactions. A consensus algorithm establishes the decentralized rules for updating this state \cite{ConsensusSurvey}, such as determining which peer is responsible for computing and communicating the next block of data. This type of blockchain technology is classified as public and permissionless, as any peer can send and receive transactions (public) and join the consensus without restrictions (permissionless). However, the consensus process between untrusted peers can be time-consuming; for example, Bitcoin generates a new block approximately every 10 minutes, resulting in low throughput. Additionally, to participate in the consensus, a peer must store the entire blockchain (hundreds of GB in size) \cite{BitcoinBC}, receive and validate all generated transactions and blocks within the network, and relay them to other peers.

To address these issues, private and permissioned blockchains were developed \cite{NeedBC}, focusing on smaller-scale systems where peers are known but not trusted. The typically smaller size of a private P2P network allows for more efficient consensus algorithms, improving throughput while maintaining data privacy within the network (private) and restricting the peers that can join the consensus (permissioned). Hyperledger Fabric \cite{HFabric} is currently the most widely used framework for building such networks.

Distributed Ledger Technology (DLT) serves as a generalization of blockchain technology, encompassing implementations that do not solely rely on a chain of blocks. For instance, IOTA utilizes a Direct Acyclic Graph known as Tangle \cite{Iota}. Public and permissionless DLTs offer high security, decentralization, and transparency but suffer from limited throughput, potentially high transaction fees, and non-compliance with privacy-preserving regulations. In contrast, private and permissioned DLTs have higher throughput and lower costs, but are less decentralized, and their security properties apply only to internal participants \cite{NeedBC}.

According to Yang et al. \cite{LedgerDB}, many private and permissioned DLT applications aim to ensure data tamper resistance, which can be achieved through cryptographic techniques that do not necessitate a decentralized architecture. Furthermore, external parties do not have any guarantees of the integrity of data disclosed by permissioned DLT participants, as they may collude and provide falsified data not present in the ledger. To address this problem, Yang et al. introduced the concept of Centralized Ledger Technology (CLT) and an implementation called LedgerDB. LedgerDB is a blockchain-like data structure that outperforms private and permissioned DLTs in terms of transactions per second, as it does not require a consensus algorithm. LedgerDB natively implements data removal to comply with privacy-preserving regulations and provides integrity proofs to external parties without access to the blockchain \cite{LedgerDB}. Other ``blockchain-based databases'' like ProvenDB \cite{ProvenDB}, BlockchainDB \cite{BlockchainDB}, and FalconDB \cite{FalconDB} also offer integrity proofs to external parties.

Certain industrial use cases present requirements that cannot be adequately addressed by standard DLT approaches \cite{LedgerDB}. Despite this, private DLT solutions are still employed in these situations \cite{Polge}, often resulting in mixed outcomes due to the technology's misuse \cite{comparative, industries}. The aim of this work is to examine the requirements of industrial use cases that pose the most significant challenges for standard DLT approaches. Based on the analysis of these requirements, we propose a novel Hybrid DLT approach that combines elements from both public and private DLT strategies.

This paper is organized as follows. Section \ref{sec:requirements} outlines a list of requirements derived from industrial use cases. Section \ref{sec:problem} discusses the difficulties in satisfying these requirements when utilizing public or private DLT methods. Section \ref{sec:solution} introduces Hybrid DLT, an innovative approach that incorporates aspects of both public and private DLTs. Section \ref{sec:discussion} examines the external auditability of Hybrid DLT and explains how each target requirement is satisfied. Section \ref{sec:variations} explores potential variations of Hybrid DLT tailored to specific use cases. Section \ref{sec:implementation} presents Traent Hybrid Blockchain, a commercial, real-time, data-intensive collaboration system designed for organizations requiring immutable data while adhering to the European General Data Protection Regulation (GDPR) \cite{GDPR}. Section \ref{sec:relatedwork} compares the Hybrid DLT approach with alternative methods. Lastly, Section \ref{sec:conclusion} offers conclusions and suggestions for future work.

\subsection{Terminology}

\noindent To improve clarity, we provide definitions for some of the terminologies used in the paper:

\begin{itemize}
    \item \textbf{Data integrity}: the property of data that has not been modified, using its original version as a reference.
    \item \textbf{Data authenticity}: the property of data that has a known source or author.
    \item \textbf{Non-repudiable data}: data whose integrity and authenticity are apparent and provable, making it impossible for the data source or author to repudiate it.
    \item \textbf{Verifiable data structure}: a data structure whose integrity can be verified using a digest.
    \item \textbf{Data history}: a data structure that identifies an ordered sequence of data. If the elements of the sequence are verifiable data structures, the history can be the list of digests.
    \item \textbf{Persistent data structure}: a data structure that always preserves the previous version of itself when it is modified.
    \item \textbf{Data consistency}: in the context of persistent data structures, the property of two time-ordered data structures where the newer data structure contains the older one as a previous version of itself.
    \item \textbf{History consistency}: a data history where data consistency holds for every pair of data structure.
    \item \textbf{Tamper-proof data storage}: storage where the integrity of the stored data is preserved by preventing any alteration.
    \item \textbf{Tamper-evident data storage}: storage where any alteration of the stored data is made apparent.
\end{itemize}

\section{Industrial use cases requirements}\label{sec:requirements}

\noindent We examined several real-world industrial use cases where data tamper resistance is a crucial feature. Two notable examples include the management of digitalized clinical trials \cite{ClinicalTrials} and a supply-chain management system capable of creating digital product passports, as envisioned by the European Commission in the Sustainable Product Regulation \cite{DigitalProductPassport}.

We have identified a list of common requirements in industrial use cases that present significant challenges when implementing DLT for achieving data tamper resistance:

\begin{itemize}
    \item \textbf{RQ1-Authentication}: all actors, particularly data producers, must be identified, and any data managed by the system must be non-repudiable by its authors.
    \item \textbf{RQ2-Privacy}: the system must be capable of restricting data access.
    \item \textbf{RQ3-Performance}: the system must be suitable for serving as the primary data layer for data-intensive, real-time applications.
    \item \textbf{RQ4-Disclosability}: the system should allow for the disclosure of persistent data (either the entire persistent data structure or a portion of it) to external actors, such as end-users, auditors, or new nodes joining the network, while maintaining the ability to prove its integrity, authenticity, and history consistency.
    \item \textbf{RQ5-Erasability}: the system must enable the erasure of portions of persistent data while preserving the ability to demonstrate the integrity, authenticity, and history consistency of the remaining data.
\end{itemize}

\section{Problem analysis}\label{sec:problem}

\noindent DLTs are a widely used form of P2P network that provides tamper-resistant storage of data. There are two main types of DLTs: public and permissionless, which use trustless consensus protocols, and private and permissioned, which rely on trusted authorities.

When evaluating only the requirements of authentication (RQ1) and privacy (RQ2), the most suitable solution would be to develop the system as a private DLT network that supports private channels, such as Hyperledger Fabric \cite{HFabric-Channels}.

The performance requirement (RQ3) is quantitative in nature. It is widely known that private DLT solutions usually offer better performance than public DLT solutions \cite{Garriga}. This is often achieved by using faster consensus protocols, albeit with weaker security or stronger trust assumptions, provided that the network is not large. The size of blocks is an important factor to consider, as it affects the block generation rate. A larger block can accommodate more transactions, but it also requires more time to propagate through the network. In public DLTs, the choice of block size not only affects performance but also security (as evidenced by the block size limit controversy in Bitcoin \cite{Blocksize}). However, the authors of BLOCKBENCH, a benchmarking framework for private DLTs, observed that increasing the block size in private blockchains does not necessarily improve DLT throughput \cite{Blockbench}. To enhance transactions per second, the performance metric most commonly employed for both public and private DLT approaches, DLTs typically set a maximum block size limit within the range of a few megabytes and focus on minimizing transaction sizes. However, when using a distributed ledger as the data layer for an application, it can be beneficial to store the data associated with application-level indivisible operations within a single transaction, as this avoids waiting for the generation and finality of multiple blocks, which can introduce significant delays. Consequently, a design centered on small transactions may not be ideal from a performance perspective when the size of the data associated to application-level indivisible operations at the application level is several orders of magnitude larger than the maximum transaction data size.

The requirement of disclosability (RQ4) relies on the external auditability of data management. Auditability refers to the ability to prove the correctness of a sequence of operations \cite{Bitfury}. In public DLTs, where there is no distinction between internal and external, auditability is inherent due to the immutability and transparency properties \cite{BlockchainACMaesa}. Conversely, private DLTs face challenges in achieving external auditability. When a piece of data from a private DLT is made public for the first time, there is no immediate way for someone outside the network to verify its consistency with the private ledger due to access limitations. Furthermore, even if full access to the private network is granted to an auditor, participants of the private network can still agree to rewrite history right before the auditor joins, effectively forking the ledger. Such a history rewriting fork has been publicly done on Ethereum after the DAO attack \cite{Dao}. On a smaller and controlled network, a fork can be done more easily and secretly \cite{consortium}. The same problem is present when a new node joins the private network. Therefore, when only considering RQ4, private DLT solutions must be discarded.

The requirement of erasability (RQ5) is not compatible with public DLTs. While it may be technically possible to organize public DLT ledger data in a way that allows for a limited form of data removal (such as organizing all data as off-chain data referred to by the chained blocks), unrestricted ledger data dissemination to unauthenticated nodes makes it practically impossible to identify the data joint-controllers to which data subjects must send data removal requests, as required by GDPR \cite{BlockchainGDPR}. Therefore, when considering RQ5 alone, a public DLT solution must be discarded.

When we look at all the requirements together, even if we focus only on the qualitative ones, we see a conflict between RQ4 and the rest. The typical private DLT approach falls short of meeting the RQ4 requirement, while the typical public DLT approach cannot satisfy the other requirements. This indicates that a new approach is needed to fulfill all the requirements at once.

It is important to note that a DLT solution, regardless of whether it is a private or public DLT, typically includes several common features that are not part of our target requirements. These features include:

\begin{itemize}
    \item \textbf{Distributed writes}: all nodes can simultaneously write new data to the ledger. To maintain a single coherent version of the ledger despite distributed writing, nodes agree to follow a consensus protocol. Adhering to a consensus protocol can be disadvantageous, as it may affect performance, throughput, or fork resistance. The consensus protocol may also impose limitations on the ability to exclude nodes from the network. Censorship resistance is unfavorable in typical industrial use cases, as the ability to block or exclude a node is necessary to address situations such as compromised access keys, malicious activities like identity theft, or other illegal activities.
    \item \textbf{Forced transparency}: data is disseminated to all nodes in the network. This is disadvantageous because when a ledger contains sensitive data, restricting data access is crucial. Although cryptography can be employed, it cannot be the sole means of ensuring security, as malicious nodes could collect encrypted sensitive data and wait for future key leaks or cryptosystem vulnerabilities.
    \item \textbf{Forced immutability}: data is made tamper-proof. This is disadvantageous because it goes directly against RQ5. Tamper-evidence would be a more valuable property for industrial use cases. To comply with regulations such as GDPR, the ability to selectively delete a portion of ledger data without compromising the tamper-evidence of the unaffected data would be ideal.
\end{itemize}

\section{Proposed solution}\label{sec:solution}

\begin{figure*}[htbp]
    \centering
    \includegraphics[width=0.8\paperwidth]{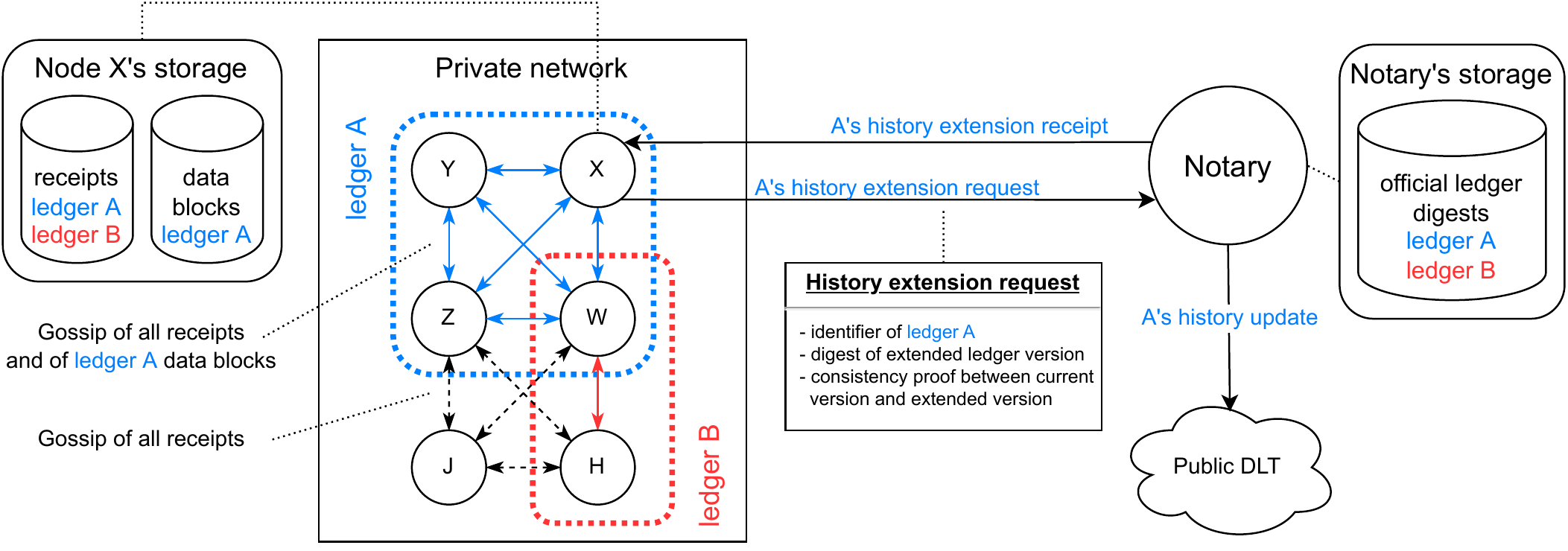}
    \caption{Hybrid DLT schematic, including a ledger extension flow initiated by a node. Data block dissemination is limited to a specific subgroup of nodes for each ledger. However, the receipts for all ledgers are distributed among all nodes.}
    \label{fig:architecture}
\end{figure*}

\noindent We propose a novel approach called Hybrid DLT that prioritizes requirements RQ1-5 while disregarding certain features such as distributed writes, forced transparency, and forced immutability, which we deemed unnecessary or detrimental. As a result, the applicability of Hybrid DLT differs from that of both public and private DLT.

Hybrid DLT is a permissioned private network where nodes create lightweight ledgers on-demand for use as private data channels between sub-groups of nodes. A Notary component maintains a unique and coherent official data history for each private ledger authoritatively, without the need to access data blocks. The Notary then publishes the data histories to a tamper-proof storage (such as a public DLT), ensuring that they are provably unique and coherent and achieving tamper-evidence for the data contained in the ledgers. Figure \ref{fig:architecture} provides an overview of the architecture.

\subsection{The components}

\noindent A \textit{node} is a peer that has been granted access to the private network of Hybrid DLT. Each node is identified by a digital certificate that specifies a public key.

\textit{Data} is stored as the payload of private ledgers. Access to these ledgers is limited to a sub-group of nodes (identified by a set of public keys) that possess permission to extend that specific ledger history.

\textit{Ledgers} are organized as an ordered list of data blocks.

Each \textit{data block} contains arbitrary data, such as a binary blob. An explicit link between consecutive data blocks is not required to be included in the block payload.

Each list of data blocks is uniquely identified (under cryptographic hashing assumptions) by a \textit{ledger digest}. For instance, one could build a Merkle tree \cite{Merkle} using the ordered data blocks as leaves and take the resulting root value as the digest.

A digest also uniquely identify the ordered sequence of data block appending operations that incrementally produced a specific list of data blocks, the data \textit{history} of the ledger.

The history, unlike the blocks, must be tamper-proof. Thus, any changes made to the block list or block contents can be detected by checking the history, making both the block list and the content of every block tamper-evident.

The \textit{Notary} is a system with authority over the official version of all ledger histories. It is identified by a digital certificate that specifies a public key.

All official ledger histories are rendered non-repudiable by the Notary, with their evolution over time tracked by \textit{receipts}, which are concise cryptographic statements generated by the Notary. The histories are demonstrably consistent, a property that can be verified using techniques such as Merkle consistency proofs \cite{rfc6962}.

\subsection{New ledger creation}

\noindent In the private network of Hybrid DLT, a new ledger is created to establish a private communication channel between a sub-group of nodes. The data blocks in the ledger contain non-repudiable messages that are disseminated exclusively among the nodes in the sub-group.

\begin{figure}[htbp]
    \includegraphics[width=\columnwidth]{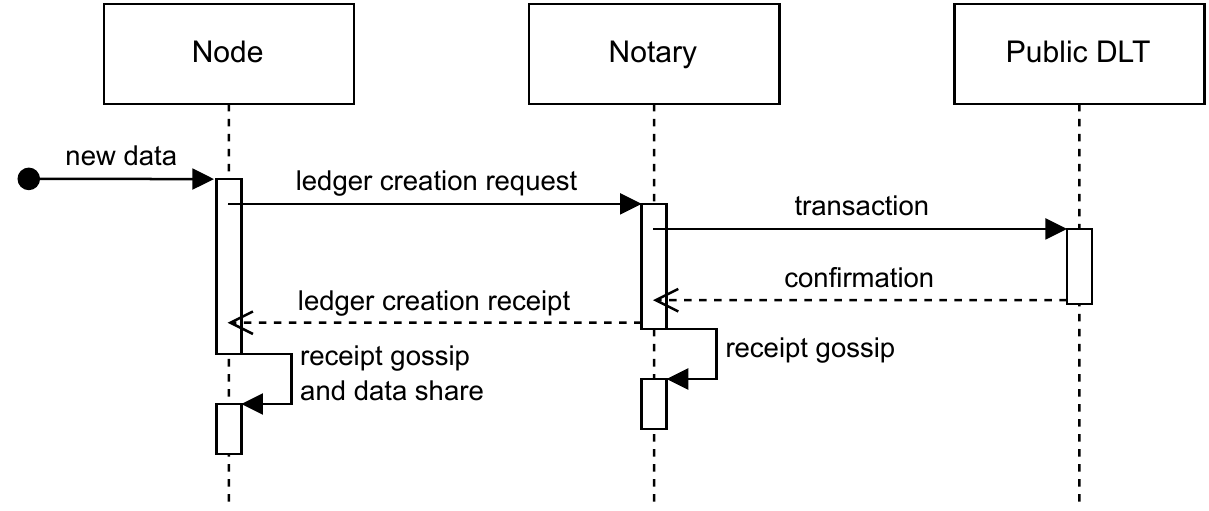}
    \caption{Sequence diagram of the new ledger creation process.}
    \label{fig:ledgercreation}
\end{figure}

To create a new ledger, one of the nodes in the sub-group initiates the process shown in figure \ref{fig:ledgercreation}. This involves choosing a new identifier for the ledger, creating a non-yet official version of the ledger, computing its digest, and constructing a \textit{ledger creation request}. The request contains the new ledger identifier, a set of authorized authors (identified by their public keys), and the initial ledger digest. Then, the node digitally signs the ledger creation request and sends it to the Notary for approval.

The Notary is responsible for verifying the request by checking the new ledger identifier's availability (it must not be already in use), the validity of the set of authors, and the signature. If the verification is successful, the Notary initializes the official history of the new ledger by associating the identifier with the initial ledger digest. The Notary stores the history in its private database and in the public DLT.

The Notary then digitally signs a \textit{ledger creation receipt}, which includes the accepted ledger creation request and useful metadata, such as a timestamp and a public DLT transaction ID. The receipt is returned to the requesting node and disseminated to the rest of the network via a gossip protocol. This receipt confirms the successful creation of the new ledger, which is now disseminated among all intended nodes of the sub-group.

\subsection{Existing ledger extension}

\begin{figure}[htbp]
    \includegraphics[width=\columnwidth]{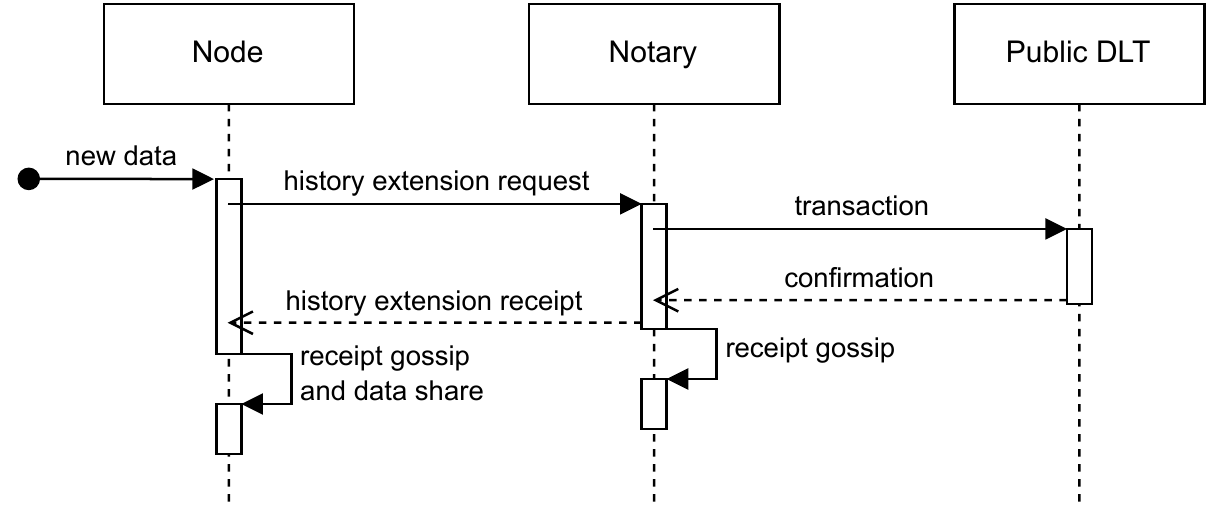}
    \caption{Sequence diagram of the existing ledger extension process.}
    \label{fig:ledgerextension}
\end{figure}

\noindent In order to append new blocks to an existing ledger, a node must initiates the process shown in Figure \ref{fig:ledgerextension}. The node must possess a secret key associated with a public key included in the set of authors of the ledger.

The authoring node first computes the extended ledger digest from the current list of data blocks extended by appending the new data blocks. To prove the consistency of the ledger's history the node also computes a consistency proof between the current official ledger digest and the extended one. For instance, when using Merkle tree roots as the ledger digests, a Merkle consistency proof \cite{Merkle, rfc6962} can be used. The node creates a \textit{ledger history extension request} containing the ledger identifier, the new digest, and the consistency proof. The request is then digitally signed and sent to the Notary for approval.

The Notary checks the request by verifying that the ledger identifier is known, the current official ledger digest matches the one used in the consistency proof, the consistency proof is valid, and the digital signature of the request can be verified by a public key included in the set of authors.

If the verification is successful, the Notary updates the ledger's digest in its private database, thus updating the ledger's history. The Notary also stores the triplet \textit{{previous ledger digest; new ledger digest; consistency proof between the previous and new ledger digests}} in the public DLT. Then the Notary creates a digitally signed \textit{ledger history extension receipt} that includes the approved request and any useful metadata, such as a timestamp and a public DLT transaction ID. The receipt is returned to the requesting node and disseminated to the rest of the network via a gossip protocol. This receipt confirms the successful extension of the ledger with the new data blocks, which are now disseminated among all the participants of the ledger-specific sub-group of nodes.

\subsection{Data disseminated inside the private network}

\noindent In the private network of Hybrid DLT, nodes exchange receipts that embed ledger creation or history extension requests, as well as data blocks. These exchanges follow two distinct policies.

\subsubsection{Receipts dissemination}

\noindent For a receipt to be considered valid, it must contain a signature that can be verified by the Notary's public key, and the request included in the receipt must have a signature that can be verified by a public key in the ledger's author list. This guarantees that the receipt cannot be repudiated by either the Notary or the request author node. The receipts are disseminated freely among all nodes in the network using a gossip protocol, ensuring that:

\begin{itemize}
    \item All nodes in the private network are aware of the official history version maintained by the Notary for every ledger.
    \item The Notary cannot maintain an inconsistent history for a ledger without making it apparent to the nodes (fork evidence).
    \item The Notary cannot maintain multiple parallel histories of the same ledger without making it apparent to the nodes.
    \item The Notary cannot accept a request whose signature can be verified by a public key not included in the ledger's author list without making it apparent to the nodes.
    \item If a node gains access to the data blocks of a previously access-restricted ledger, they can check if the blocks are consistent with the known ledger history.
\end{itemize}

If the Notary violates any of the creation or extension protocol rules, a node can detect the misbehavior thanks to the disseminated receipts. The misbehavior can be cryptographically proven to external actors. This control system reduces the level of trust that nodes need to place in Notary's authority and encourages nodes to participate in receipt dissemination.

It is important to highlight that the size of a ledger history extension receipt does not directly depend on the number or size of the appended data blocks; instead, it is related to the length of the history. For instance, when employing Merkle consistency proofs, the receipt size scales logarithmically with respect to the block list's length \cite{rfc6962}.

\subsubsection{Data blocks dissemination}

\noindent Nodes create data blocks independently and never share them with the Notary or any node without data access permission. Once a node receives a receipt from the Notary when creating or extending the history of a ledger, it can choose to follow a data sharing policy agreed upon with a sub-group of nodes in the network. One default policy is to share the data blocks exclusively with nodes in the sub-group identified by the ledger's author list keys.

If a node maintains a local replica of a ledger and becomes aware of a new history extension through receipt dissemination, it can request the recently appended data blocks from the authoring node (whose identity is discernible from the request embedded in the receipt) or another node with data access, in accordance with the implemented policy. Upon receiving new data blocks, a node can verify their consistency with the official history. If the node receives only the digest of a data block instead of its content, it can still validate it as an explicit omission (this may occur, for example, when complying with a GDPR data removal request).

When sharing data blocks, a node can send them along with relevant receipts that prove their consistency with the official history. This allows any receiver to check for inconsistencies by comparing them to receipts already in their possession. A node can share a selection of data blocks with any node in the network without previous data access to the ledger by pairing them with the relevant receipts and inclusion proofs (such as Merkle inclusion proof \cite{rfc6962}). These proofs can be generated autonomously from the data block digests or received from another node during a previous data share. The receipts and inclusion proofs are sufficient to prove that the shared data blocks are included in the portion of the history extended by a specific ledger history extension, which can then be validated by the receiving node against the known official ledger history.

\subsection{Data disseminated outside of the private network}\label{sec:solution:data}

\noindent The official ledger histories are disseminated outside the private network as multiple sequences (one for each ledger) of ledger digests, with each digest paired with a consistency proof to the previous one. The Notary writes these sequences to an external public DLT, using its official address, ensuring they cannot be tampered with or repudiated.

Nodes expect the Notary to maintain the official histories within the private network in sync with those maintained externally. To verify this, every node can compare all receipts disseminated within the private network with the histories published on the public DLT.

The provably consistent official histories stored in the public DLT allow every actor, even those external to the private network, to independently verify that all the data written by the Notary's official address describes a unique coherent history for every ledger identifier, ensuring external public auditability.

\begin{figure}[htbp]
    \includegraphics[width=\columnwidth]{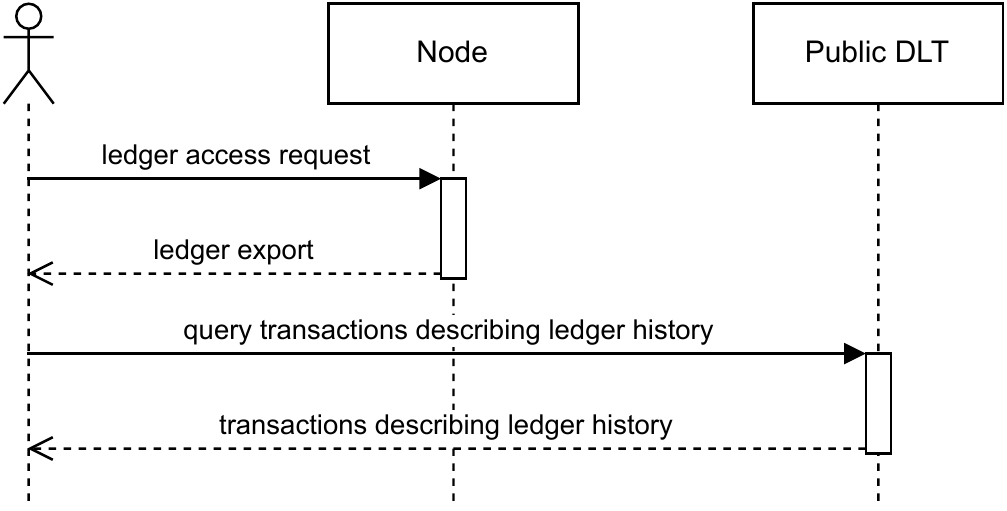}
    \caption{Sequence diagram of the ledger export and the audit of its history.}
    \label{fig:audit}
\end{figure}

An interesting scenario occurs when a node shares a private ledger with an external actor through an export (either the entire ledger or just selected data blocks paired with inclusion proofs). The external actor can audit the Notary's activity and retrieve the data history of the exported ledger from the public DLT, verifying the data integrity of the export against the tamper-proof history. Figure \ref{fig:audit} illustrates this scenario.

\section{Discussion}\label{sec:discussion}

\noindent In this section, we compare the proposed Hybrid DLT approach with private and public DLTs.

To begin, let us view the private DLT approach as a public DLT with the following variations:

\begin{itemize}
    \item Access to the network is restricted and can be revoked, thereby losing censorship resistance.
    \item Nodes are identified, and transactions are authenticated, thereby losing anonymity or pseudonymity.
    \item External actors without access to the ledger cannot audit, thereby losing public auditability.
    \item Nodes in the network can collude to modify data \cite{consortium}, thereby losing public tamper-proofness.
\end{itemize}

On the other hand, private DLT introduces a data control mechanism (by disseminating data only within the private network) and an improvement in performance \cite{consortium, comparative}. This trade-off in properties might be perceived as a downgrade; however, it represents a effective approach to adapt the technology to a different category of use cases.

Similarly, the Hybrid DLT approach can be seen as a modification of the private DLT approach, where:

\begin{itemize}
    \item Nodes still create new data, but they need to obtain authorization in the form of a receipt from the Notary to extend the official ledger history, thereby losing distribution of writes.
    \item Ledgers data blocks are disseminated only among a subset of the nodes in the network, thereby losing forced transparency.
    \item Some data blocks can be deleted and forgotten without affecting the ability to prove the integrity and authenticity of the remaining ones, thereby losing forced immutability.
\end{itemize}

The Hybrid DLT approach provides more control over data block dissemination among nodes in the network, allowing for ledger-specific data access policies. The transactions issued by the Notary to the public DLT ensure public auditability of official ledger histories. Thanks to this feature, any external actor who receives one or more data blocks of a ledger can verify its integrity and consistency through the public history, making the data blocks tamper-evident. The removed properties and the new ones allow for an additional shift in the area of applicability, making the Hybrid DLT suitable for certain use cases where neither public nor private DLT approaches can meet all requirements.

\subsection{Auditability}

\noindent The Notary's actions outside of the private network involve writing to the public DLT to maintain a single, official, and coherent public history for each ledger. The ledger histories can be publicly audited by executing the following checks against the data written in the public DLT by the official Notary's address:

\begin{itemize}
    \item Ensure that for every ledger identifier found, there is only one initialization transaction associating the identifier with a specific initial ledger digest value.
    \item Ensure that for every ledger identifier found, all the history extension transactions associated with that identifier:
    \begin{itemize}
        \item Are valid, i.e. the contained consistency proof is verified.
        \item Express a coherent history, i.e. every proof proves consistency between the previous ledger digest value and the new one.
    \end{itemize}
\end{itemize}

Thanks to the dissemination of the receipts, nodes can internally audit that the public and private histories are in sync, as described in Section \ref{sec:solution:data}. In case of Notary misbehavior, nodes can create a proof of misbehavior, and the inclusion of receipts enables external actors to verify this proof even if the original audit was internal. This is an additional control system, whose effect is to reduce the level of trust that nodes need to place in the Notary.

Public auditability enables the validation of the integrity and proper management of any ledger history, including the period preceding the auditor's first knowledge of that ledger. This allows new nodes to join the private network without relying on trust in the existing nodes.

\subsection{Requirements satisfaction}

\noindent In this section, we will explain how the proposed solution meets all the target requirements.

\subsubsection{RQ1-Authentication}

\noindent The main component of the proposed solution is a private and permissioned network, where any action creating or extending a private ledger is required to be digitally signed by both the author and the Notary authority.

\subsubsection{RQ2-Privacy}

\noindent The ledgers that are created and extended within the private network can only be accessed by a subset of nodes that agree to adopt a policy limiting the dissemination of data blocks. Additionally, the data blocks are not transmitted to the Notary, and the data written in the public DLT is just an indirection of ledger data.

\subsubsection{RQ3-Performance}

\noindent The proposed solution has a Notary that accepts creation and extension requests without knowing the size or format of the appended data blocks. This means that the throughput of the receipt system is not affected by the size of the data in the private ledgers.

The interaction between nodes and the Notary is a client-server system, where the Notary prevents forks and is audited using disseminated receipts, obviating the need for a consensus protocol. The nodes can store and transmit the data blocks using standard high-performance solutions since the integrity and authenticity of the data blocks can be verified using just the receipts.

The absence of a single global ledger and the ability to segregate data into contextual ledgers \cite{channels} (e.g., to store data generated during the execution of a specific business process) that can be created immediately as needed, allows nodes to spend resources only on activities relevant to their interests. Furthermore, the network and computational resources required to participate in a specific ledger are proportional to its level of activity (data block appends), reducing the cost of inactive ledgers to storage-only.

\subsubsection{RQ4-Disclosability}

\noindent Selective disclosability is achieved through the following process: a node that has access to the data blocks and receipts of a ledger can create an archive that includes a selection of data blocks, associated receipts, and inclusion proofs (such as Merkle inclusion proofs \cite{rfc6962}). These proofs can be generated autonomously by the node, and demonstrate that the data blocks are part of the specific portion of the history extended by an official ledger history extension. Any consumer of the archive can verify the authenticity and integrity of the data by examining the proofs and the official ledger history available on the public DLT. Additionally, by auditing all the public DLT transactions written by the official Notary address, any consumer can verify that the Notary has maintained a single consistent history of the disclosed ledger.

\subsubsection{RQ5-Erasability}

\noindent Nodes can delete and replace a specific data block with its digest without compromising the integrity and authenticity of the ledger structure or non-repudiability. This is thanks to the separation of data (contained in the data blocks) and history (contained in the receipts, which also include digital signatures of the authors of the deleted blocks).

\section{Variations}\label{sec:variations}

\noindent This section will discuss three variations that have been explored for the described solution.

\subsection{Notary as a trusted repository for data blocks}

\noindent In this variation, nodes send data blocks along with their ledger creation and extension requests to the Notary, entrusting it with data access. However, this approach goes against RQ2-Privacy, even if the data blocks are encrypted. This variation is only appropriate for scenarios where the Notary operator is trusted by all nodes.

When the Notary acts as a trusted repository for data blocks, it provides several advantages:

\begin{itemize}
    \item The Notary can grant access to specific data blocks to nodes that can prove their access rights, without requiring nodes to disseminate the blocks. This prevents the case of a node authoring blocks but refusing to share them with others.
    \item A safe repository for data recovery exists in case of data loss in local node storage.
    \item The Notary can certify, in addition to producing a receipt, that a list of data blocks coherent with the extended ledger digest existed at a given time.
\end{itemize}

\subsection{Notary as a policy compliance enforcer}

\noindent When using the variation of \textit{Notary as a trusted repository for data blocks}, an additional useful feature can be implemented. If the first block of the ledger includes, in plaintext, a ledger-specific policy that sets specific requirements (such as maximum data block size, allowed plaintext structures, or minimum number of authors required to sign an extension request), the Notary can enforce it when determining whether to approve or reject history extension requests. This adds new semantic implications to the issuance of a receipt.

\subsection{Delayed notarization}

\noindent Our proposed approach requires the Notary to store data in a public DLT before issuing a receipt for a ledger history extension request. However, this interaction can cause delays that may conflict with the use case requirements, depending on the specific public DLT used.

To address this issue, we suggest a variation where the Notary returns the receipt before creating the public DLT transaction. The transaction can then be created immediately after, or delayed and scheduled differently, such as once a day for each ledger. If multiple history extensions are accepted between two consecutive notarization events, the consistency proof stored in the public DLT must prove the consistency between the ledger digest values at the time of the first and second notarization.

This approach allows for higher throughput of ledger extensions, but it introduces a time window of eventual desynchronization between the official histories known internally and externally to the private network.

\section{A commercial implementation of Hybrid DLT}\label{sec:implementation}

\noindent The Hybrid DLT solution design and requirements were derived from analyzing real-world industrial use cases rather than hypothetical scenarios, such as management of digitalized clinical trials and supply-chain management system capable of creating digital product passports.

To implement the Hybrid DLT, Traent Hybrid Blockchain (\url{https://traent.com/}), a commercial product, was developed as a collaboration system for organizations. Traent provides a service that verifies and stores the identity data of individuals and organizations in private ledgers, whose access is limited based on the preferences of the organizations and individuals.

Organizations manage independently their own nodes, and members of specific organizations operate these nodes through authenticated access to the organization nodes API.

To interact with the Hybrid DLT system, organization members can use a user-friendly frontend web application hosted directly by the nodes. The application allows them to easily produce and consume data, create new private ledgers, and share them with members of other organizations. Shared ledgers are replicated among involved organizations and can be used as private collaboration channels for exchanging data.

The Notary, which publishes the official ledger histories on Algorand \cite{Algorand}, is operated by Traent.

Traent Hybrid Blockchain has been fully implemented and deployed in the Amazon Web Services cloud, and multiple usage scenarios have been tested, including:

\begin{itemize}
    \item 2/5/20 organizations, with up to 10 organizations sharing the same ledger.
    \item Maximum block size of 1 MB, 30 MB, 100 MB, or 1 GB.
    \item Writing data generated from IoT sensors at a frequency of 100 transactions every second (using the \textit{Delayed notarization} variation).
    \item Notarization with immediate or delayed frequency (1 hour or 24 hours).
    \item Stress tests on the Notary:
    \begin{itemize}
        \item Writing every second 300 blocks of 1 MB on a single ledger.
        \item Writing every second 1000 blocks of 1 MB on 100 different ledgers.
    \end{itemize}
\end{itemize}

The tests were performed using the variations \textit{Delayed notarization}, \textit{Notary as a trusted repository for data blocks}, and \textit{Notary as a policy compliance enforcer}. The Notary stress tests were run on an Intel Core i9 10900 with 10 cores, 32 GB of RAM DDR4, and 1TB SSD.

It was experimentally verified that both the Notary and a node can safely run on a t3.small Amazon EC2 T3 Instance (Intel Xeon Platinum 8000 series processor with a clock speed of up to 3.1 GHz, 2 vCPUs, 2 GB of RAM), with an observed delay for data synchronization after a node executes a ledger extension smaller than 0.1 seconds.

The workload of a single node is solely determined by the operations affecting the ledgers shared with it, without being impacted by the activity level of the entire network. This feature allows for efficient horizontal scaling, and means that the storage, CPU, and RAM requirements for nodes can vary widely.

\section{Related work and comparison}\label{sec:relatedwork}

\noindent When comparing the proposed Hybrid DLT approach to other DLTs, a major challenge is measuring performance. Transaction throughput, which measures the number of transactions per second, is commonly used as the main performance indicator for both public and private DLT approaches. However, for data-intensive real-time applications, metrics such as data throughput (bits per second) and total single transaction latency \cite{hyper-blocksize} are more relevant.

LedgerDB \cite{LedgerDB} was created to address some of the requirements identified in this study. It meets requirements such as RQ1-Authentication, RQ3-Performance, RQ4-Disclosability, and RQ5-Erasability by using a centralized ``blockchain-based database'' service. However, it does not satisfy the RQ2-Privacy requirement of data access limitation. For cases where data access can be granted to a trusted third-party, comparing LedgerDB and the Hybrid DLT approach with the \textit{Notary as a trusted repository for data blocks} variation (see Section \ref{sec:variations}) is particularly interesting. One significant difference is that the Hybrid DLT approach achieves external auditability through technological means, using a trustless public DLT as history consistency storage, while LedgerDB relies on a trusted third-party timestamping authority. Additionally, the role of the centralized component is more limited in the Hybrid DLT approach: with signed receipts and data written to the public DLT, a node can independently prove the integrity and authenticity of a given ledger state to external actors without requiring the cooperation of the central entity. This makes the system more robust and preserves data access and provability of data properties even if the central authority is unavailable.

We believe that LedgerDB can be used as the data layer for the Notary in a Hybrid DLT solution if the Notary stores the necessary data for an external audit on a public DLT and produces suitable receipts when an extension request is accepted.

Another approach worth considering is Hyperledger Fabric \cite{HFabric}, a framework developed by the Hyperledger project of the Linux Foundation to establish and operate private DLTs. With Hyperledger Fabric, it is possible to create multiple private channels on the same DLT network between various subsets of nodes \cite{HFabric-Channels}. Each private channel is an independent ledger with customizable policies and data dissemination restricted to channel participants. To create and append blocks to the ledger, a channel requires an orderer service that collects all transactions sent by channel participants, orders and packs them into new blocks. When Hyperledger Fabric is configured to only use private data collections \cite{HFabric-Collections} (databases that store the state of the channel, replicated privately only among the participating nodes, whose content is referred in the ledger blocks by hash), it satisfies the requirements of RQ1-Authentication, RQ2-Privacy, and RQ5-Erasability.

We believe that Hyperledger Fabric, when configured to only use private channels and private data collections, can be employed as the private network component of a Hybrid DLT implementation. However, some requirements need to be addressed:

\begin{itemize}
    \item RQ3-Performance: to support real-time applications, the orderer service should not introduce any delay waiting for additional transactions to be included in a new block. Instead, it should produce a new block for every transaction. This would reduce the total single transaction latency, but would also decrease the total transaction throughput \cite{hyper-throughput, hyper-blocksize}. Furthermore, Hyperledger Fabric requires the orderer to verify that each proposed transaction has been signed by a policy-defined number of endorsing nodes; if this number is greater than one, it would increase total single transaction latency \cite{hyper-blocksize}.
    \item RQ4-Disclosability: although Hyperledger Fabric is not externally auditable per se, the system described in this work can make the ledgers of private channels publicly auditable by storing a tamper-proof stream of digests and corresponding consistency proofs to a public DLT. The orderer will need the same authority as the Notary described in this work.
\end{itemize}

\section{Conclusion}\label{sec:conclusion}

\noindent We have identified certain industrial use cases (e.g., digitalized clinical trials, supply-chain management system capable of creating digital product passports) that require some properties of private and public DLTs, while others may be detrimental. To address this, we propose the Hybrid DLT approach, which combines the advantages of both while shifting to a different area of applicability.

In the proposed approach, nodes within a private network collaboratively create and extend private ledgers with limited data block dissemination. An authoritative component called the Notary maintains a single official coherent history for each private ledger without requiring access to data blocks, making it tamper-proof using a public DLT solution, and providing tamper-evidence on ledger data.

The Notary's external activity is publicly auditable, while nodes within the network can fully audit the Notary's internal activity. In case of misbehavior, nodes can create a proof that external actors can verify as evidence of the misbehavior.

We have deployed a specific commercial implementation of the Hybrid DLT approach. Private ledgers serve as the main data layer for an application, supporting real-time latency even in data-intensive scenarios without requiring high-performance hardware.

\subsection{Future work}

\noindent Our proposed solution for ledger management involves the Notary writing to a tamper-proof storage a sequence of digests and consistency proofs for each ledger. In Section \ref{sec:variations}, we discussed a variation called \textit{Delayed notarization} that reduces the number of notarization operations for frequently updated ledgers. We are now exploring a new approach (that will be presented as a separate work) that involves storing in the public DLT only the digests of a new data structure derived from all the ledger histories. This would enables us to provide verifiable proofs of history uniqueness and consistency for every possible ledger using a single sequence of digests, making the number of notarization operations independent of the number of maintained ledger histories.

Additionally, we are exploring methods to enhance auditability and further reduce the level of trust that nodes need to place in the Notary. This can be achieved by incorporating independent monitoring systems at various points in the infrastructure.

We are also considering scenarios where multiple nodes send frequent requests for history extensions for the same ledger. When several requests are received simultaneously, accepting one modifies the official current ledger digest and invalidates the consistency proofs in all the others, requiring the authoring nodes to rebuild the requests using the new official ledger digest. To improve throughput and minimize the rejection rate, we are exploring an approach that permits nodes to authorize the Notary, within certain constraints, to reorder their history extensions before accepting them.

\section{Disclaimer}

\noindent This research is sponsored by Traent and the proposed solution, as well as the implementations using the variations and approaches described in the \textit{Future work} section, are patents pending by Traent.

\printbibliography

\end{document}